\documentstyle[aps,prd,twocolumn,epsf]{revtex}
\tighten
\newcommand \beq{\begin{eqnarray}}
\newcommand \eeq{\end{eqnarray}}

\begin{document}
\draft
\title{Kaon Condensation in Dense Matter}

\author{J. Carlson$^1$,
H. Heiselberg$^2$ and V. R. Pandharipande$^3$}

\address{
1) Theoretical Division, Los Alamos National Laboratory,
Los Alamos, New Mexico 87545, USA\\
2) NORDITA, Blegdamsvej 17, DK-2100 Copenhagen \O., Denmark\\
3) Dept. of Physics, Univ. of Illinois at Urbana-Champaign,
   1110 West Green St., Urbana, Illinois 61801-3080, USA  
}

\date{\today}
\maketitle
\begin{abstract}
The kaon energy in neutron matter is calculated analytically with the
Klein-Gordon equation, by making a Wigner-Seitz cell approximation and
employing a $K^-N$ square well potential.  The transition from the low
density Lenz potential, proportional to scattering length, to the high
density Hartree potential is found to begin at fairly low densities.
Exact non-relativistic calculations of the kaon energy in a simple
cubic crystal of neutrons are used to test the Wigner-Seitz and the
Ericson-Ericson approximation methods.  All the calculations indicate
that by $\sim 4$ times nuclear matter density the Hartree limit is
reached, and as the Hartree potential is less attractive, density for
kaon condensation appears to higher than previously estimated.
Effects of a hypothetical repulsive core in the $K^-N$ potential are
also studied.
\end{abstract}

\vspace{1.5cm}

 Kaon condensation in dense matter was suggested by Kaplan and Nelson
\cite{KN}, and has been discussed in many recent publications
\cite{BLRT,PPT,Weise}. Due to the attraction between $K^-$ and
nucleons its energy decreases with increasing density, and eventually
if it drops below the electron chemical potential in neutron star matter in
$\beta$-equilibrium, a Bose condensate of $K^-$ will appear.

The kaon-nucleon interaction in vacuo have been described by Brown,
Lee, Rho, and Thorsson \cite{BLRT} by an effective Lagrangian based on
chiral perturbation theory. The $K^-n$ interaction is well described
and is fortunately not affected much by resonances as is the $K^-p$
interaction. Kaiser, Rho, Waas and Weise
\cite{Weise} have used energy dependent $\bar{K}N$ amplitudes
calculated in a coupled-channel scheme starting from the chiral SU(3)
effective Lagrangian. They correct for correlation effects in
a nuclear medium and find that $K^-$'s condense at densities above
$\sim 4\rho_0$, where $\rho_0=0.16$ fm$^{-3}$ is normal nuclear
matter density. This is to be compared to the central density of
$\sim4\rho_0$ for a neutron star of mass 1.4$M_\odot$ according to the
estimates of Wiringa, Fiks and Fabrocini \cite{WFF} using realistic
models of nuclear forces. The condensate could change the structure
and affect maximum masses and cooling rates of massive neutron stars.

In this letter we calculate the kaon energy in neutron matter using 
the Wigner-Seitz approximation for the Klein-Gordon equation 
of kaons in neutron matter.  Our formulation is exact in both, the 
low density and the high density limits. 

We assume that the kaon-nucleon interaction is via the 
Weinberg-Tomozawa vector potential $V(r)$. 
In the analysis of Ref. \cite{Speth} the $K^+N$ interaction was also
found to be dominated by $\omega$ and $\rho$ vector mesons.
The energy of the kaon-nucleon center-of-mass system with respect to
the nucleon mass is then
\beq
   \omega = \sqrt{k^2+m_K^2}+V(r)+\frac{k^2}{2m_N}  \,, \label{E}
\eeq
where $m_N=939.5$~MeV is the neutron mass, $m_K=494$~MeV the kaon
mass, and $k$ is the kaon momentum (we use units in which $\hbar$ and
$c$ are unity) in center-of-mass frame.
We have included the recoil kinetic energy of the nucleon assuming that 
terms of order $k^4/8m_N^3$ and higher can be neglected.
For a relativistic description of the kaon in a vector potential we
employ the following recoil corrected Klein-Gordon (RCKG) equation 
obtained by quantizing Eq. (\ref{E}) ($k=-i\nabla$)
\beq
   \left\{ (\omega-V(r))^2 +\frac{m_N+\omega-V(r)}{m_N}\nabla^2-m_K^2 \right\} 
     \phi = 0       \,.\label{RCKG}
\eeq
The shape of the kaon-nucleon potential is not known.  
In most of our work we approximate it with a square well:
\beq
   V(r)= -V_0\Theta(R-r) \,.    \label{VSW} 
\eeq
Yukawa and repulsive core shaped potentials are less favorable for 
kaon condensation. 
The range of the interaction $R$ and the potential depth $V_0$ are
related through the s-wave scattering length:
\beq
   a=R-\frac{\tan(\kappa_0R)}{\kappa_0}\,,  \label{a}
\eeq
where $\kappa_0^2=(2m_KV_0+V_0^2)m_N/(m_N+m_K+V_0)$.
If $V_0\ll m_K$, this reduces to the nonrelativistic result 
$\kappa_0^2=2m_RV_0$, where
$m_R=m_Km_N/(m_K+m_N)$ is the kaon-nucleon reduced mass. 
The $K^-n$ scattering length is negative corresponding to 
a positive $V_0$.

The kaon-nucleon scattering lengths are, to leading order in chiral
meson-baryon perturbation theory, given by the Weinberg-Tomozawa 
vector term 
\beq
   a_{K^\pm n}\,=\, a_{K^\pm p}/2
    \,=\, \pm\frac{m_R}{8\pi f^2}\,\simeq\, \pm 0.31 {\rm fm} \,, \label{Born}
\eeq
where $f\simeq90$ MeV is the pion decay constant. Scalar and other higher order
terms have been estimated and we shall use the value  $a_{K^-n}=-0.41$ fm
from \cite{BLRT}. This estimate does not include the $\Sigma\pi$ decay
channels which account for the complex parts of the 
$K^-N$ scattering lengths \cite{LJMR,Weise2}.
The empirical scattering lengths extracted 
from scattering measurements as well as kaonic atoms are
$a_{K^-n}=(-0.37-i0.57)$ fm and $a_{K^-p}=(0.67-i0.63)$ fm.

Effective Lagrangians predict $a_{K^-p} \simeq -0.82$ fm, 
in perturbation theory. However, it
becomes positive due to the presence of the nonperturbative 
$\Lambda(1405)$ resonance. Interestingly, 
we can estimate the range of our square well potential assuming
that the $\Lambda(1405)$ is a $K^-p$ bound state. 
As in \cite{Weise2} we assume the $K^-p$ potential to be twice as
strong as the $K^-n$ potential given by Eq. (\ref{a}).
In order to form a $K^-p$ bound state with binding energy of 
$m_p+m_K-m_\Lambda(1405)=27$ MeV, and have $a_{K^-n}=-0.41$ fm 
we need $R\simeq0.7$ fm. 
For this reason we present results for 
$R=0.7$ fm; however our main conclusions are valid for reasonable 
values of $R$.  
Using $a=-0.41$ fm and $R=0.7$ fm we obtain $V_0$=122 MeV 
with RCKG and =126 MeV with the non-relativistic. 
It is much weaker than the $NN$ interaction, and has a small 
relativistic correction.

In neutron matter at low densities the kaon energy deviates from its 
rest mass by the ``Lenz'' potential proportional to the scattering
length
\beq
   \omega_{Lenz} - m_K = \frac{2\pi}{m_R}\, a_{K^-n}\,\rho
               \,,\label{Lenz}
\eeq
which is the optical potential obtained in the impulse 
approximation.
At high densities the kaon energy deviates from its rest mass 
by the Hartree potential:
\beq
   \omega_{Hartree} - m_K = \rho \int V_{K^-n}(r) d^3r = - \frac{4\pi}{3}R^3 \rho V_0 
                    \,,\label{Hartree}
\eeq
which, as shown in \cite{PPT}, is considerably less attractive.

To estimate the transition from the low density Lenz potential to
the high density Hartree potential we solve the Klein-Gordon equation
for kaons in neutron matter in the Wigner-Seitz (WS) cell approximation.
It allows an analytical calculations of
the kaon energy which will be compared to numerical calculations on a
cubic lattice. The cell boundary condition contains the important scale 
of the range of kaon-neutron correlations 
in matter, and 
gives the correct low density (Lenz) and high density (Hartree) limits.

The RCKG equation for s-waves is
\beq
  \left\{ \left(\frac{m_N+\omega-V}{m_N}\right) \frac{d^2}{dr^2}
         +(\omega-V)^2 - m_K^2\right\} r\phi = 0 \,,\label{KG1}
\eeq
which has the solution
\beq
  u \equiv r\phi = \left\{ \begin{array}{lll}
\sin(\kappa r)  \,,&\mbox{for $r\le R$  }\\
Ae^{kr}+Be^{-kr}\,,&\mbox{for $r\ge R$  }
            \end{array} \right\} \,.\label{u}
\eeq
where $\kappa^2=((\omega+V_0)^2-m_K^2)m_N/(m_N+\omega+V_0)$
and $k^2= (m_K^2-\omega^2)m_N/(m_N+\omega)$. 
By matching the wave function $u$ and its derivative at $r=R$ we obtain
\beq
   \frac{k}{\kappa}\tan(\kappa R) = 
    \frac{e^{2kR}+\frac{B}{A}}{e^{2kR}-\frac{B}{A}} 
          \,.\label{1}
\eeq

\begin{figure}
\epsfxsize=8.6truecm 
\epsfbox{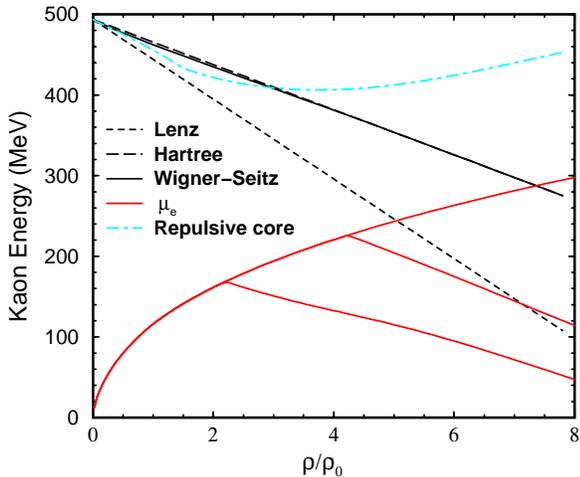}
\caption[]{Kaon energy as function of neutron density. 
The results obtained with the WS method (Eq.12) 
are shown by full curve. 
At low densities they approach the Lenz result
(Eq. (\ref{Lenz}), short-dashed curve) and at high densities they
approach the Hartree result (Eq.(\ref{Hartree}), long-dashed curves),
while the dash-dotted curve gives the WS energies for 
the repulsive core potential.
The electron chemical potential $\mu_e(\rho)$ is shown (lower full curves) 
for the equation of state for beta stable nuclear matter
\protect\cite{APR98,physrep} without and with
a transition to a mixed phase of quark matter for bag constants 
$B=122$ and $B=200$ MeV fm$^{-3}$. Quark matter admixtures reduce 
the $\mu_e$ as shown.}
\label{fig1}
\end{figure}

In the WS approximation $\phi'(r_0)=0$, where the cell size $r_0$
is given by the density $\rho=(4\pi r_0^3/3)^{-1}$. From Eq. (\ref{u})
this implies
\beq
   kr_0 = \frac{e^{2kr_0}+\frac{B}{A}}{e^{2kr_0}-\frac{B}{A}} 
            \,. \label{2}
\eeq
Eliminating the coefficient $B/A$ from Eqs. (\ref{1}) and (\ref{2}) gives
\beq
   \frac{k}{\kappa}\tan(\kappa R)=\frac{e^{2k(R-r_0)}-(1-kr_0)/(1+kr_0)}
         {e^{2k(R-r_0)}+(1-kr_0)/(1+kr_0)} \,, \label{3}
\eeq 
which determines $k$ and thus the kaon energy. The resulting kaon energy is
shown in Fig. (\ref{fig1}).

At low densities, $r_0\gg R$, the kaon energy is $\omega\simeq m_K$ and
$k\sim0$. Since also $kr_0\ll1$, we can expand the r.h.s. of Eq. (\ref{3})
and find
\beq
   \frac{\tan(\kappa R)}{\kappa} = R + \frac{1}{3} k^2r_0^3 
    + \frac{1}{5} k^4r_0^5 + ...\,. \label{4}
\eeq
Now we can extract the kaon energy
\beq
   \omega^2 &-& m_K^2 \,=\, - k^2 \frac{m_K}{m_R} \nonumber\\
     &=& \frac{m_K}{m_R} 4 \pi a_{K^-n} \rho 
        \left( 1+\frac{9}{5}a_{K^-n}(\frac{4\pi}{3}\rho)^{1/3}+...\right) 
   \,. \label{Lenz2}
\eeq
The linear part of Eq. (\ref{Lenz2}) is the Lenz potential, Eq. (\ref{Lenz}),
since $\omega + m_K \sim 2 m_K $ at small $\rho$.
The next order scales with $\rho^{4/3}$ and 
it becomes a quarter of the leading Lenz potential at 
$\rho\simeq\rho_0/16$. This demonstrates that the Lenz potential 
of kaons is valid only at very small densities, and is of 
limited interest for kaon condensation.

At the density where $r_0=R$ equations (8) and (9) are solved by 
a constant $\phi$ implying $\omega=m_K-V_0$, the Hartree 
energy. At even higher densities the
two-body potentials overlap with each other, and the Hartree approximation
presumably becomes valid. 
It gives for $r_0\le R$:
\beq
   \omega = m_K - V_0\left(\frac{R}{r_0}\right)^3 \,.\label{Hartree2}
\eeq
Note that this equation is valid in both relativistic and non-relativistic 
mean field limits; only the value of $V_0$ is influenced by relativistic 
effects in the scattering process.
The energies obtained with the Lenz and Hartree approximations
are also shown in Fig. (\ref{fig1}).  
The cross-over from the Lenz to the Hartree limit takes place at rather 
small densities. 

If the kaon-nucleon potential has a short range repulsive core of 
radius $R_c$, a stronger attractive potential $-V_0$ 
is needed at $R>r>R_c$ to obtain
the same scattering length. In Fig. (\ref{fig1}) we show an example of the
effect of a repulsive core. The repulsion is chosen such that the
interaction has zero volume integral, i.e., the core potential is
$V_c=V_0(R^3/R_c^3-1)$. We choose
$R=1$~fm and $R_c=R/2$.  
In order to obtain the scattering length $a_{K^-n}=-0.41$~fm, with 
non-relativistic kinematics, the
attractive potential depth is $V_0=153$MeV.  At very low densities the
kaon energy calculated with the WS approximation, 
follows the Lenz potential and is not affected by the
presence of a repulsive core.  At intermediate densities, $r_0\sim R$,
it is actually lower with than without a repulsive core.
However, at higher densities the kaon energy approaches $m_K$, 
the Hartree limit for an interaction with zero volume integral. 
The presence of a repulsive core will
thus further reduce the possibility of kaon condensation.

\begin{figure}
\epsfxsize=8.6truecm 
\epsfbox{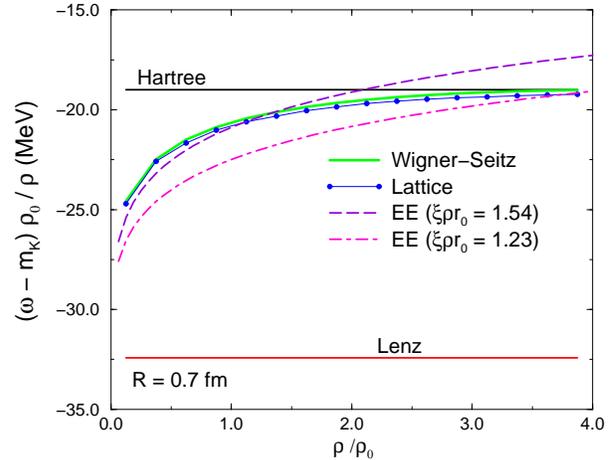}
\caption[]{Comparison of Wigner-Seitz energies, scaled by 
the factor $\rho_0/\rho$, with numerical exact results 
for a simple cubic crystal (dots), and Ericson-Ericson energies for 
$\xi \rho r_0 = 1.54$ and 1.23. 
These non-relativistic calculations use static,
infinitely massive nucleons.}
\label{fig2}
\end{figure}

In order to test the WS approximation we 
consider an extreme limit in which the nucleons are infinitely massive
and fixed in a simple cubic lattice.  Using non-relativistic kinematics, it is
simple to solve for the ground state of the kaon in this
lattice. We evaluate the imaginary-time propagator $\exp[-H\tau]$
for small $\tau$ on a grid, and iterate $\Psi(\tau +\Delta\tau)
=\exp[-H\tau]\Psi(\tau)$ until convergence to the ground state.  This
can be done efficiently by starting with a course grid and then using
a finer mesh as the iterations proceed.
The WS results for 
non-relativistic kinematics and infinitely heavy nucleons are 
compared with the lattice results for a $R=0.7$ fm square well 
potential with $a_{K^-n}=-0.41$ fm.  The structure 
of the lattice is not important; results for the bcc lattice, for example,
fall between the WS and the exact results for the simple cubic lattice.
In Fig. (\ref{fig2}) we show $(\omega_K-m_k)\rho_0/\rho$ 
which equals $2\pi a_{K^-n} \rho_0 /m_R$ in the Lenz and 
$-V_0R^34\pi\rho_0/3$ in the Hartree limits. 
We find little difference between the two sets of
results throughout the range of densities considered.

In Ref. \cite{Weise} the kaon energy is corrected for correlations 
in the medium.  This effect is analogous to
the Ericson-Ericson (EE) correction for pions in a nuclear medium \cite{EE}
and the Lorentz-Lorenz effect in dielectric medium \cite{BB}. 
It is interesting to compare the kaon energy $\omega_{EE}$ obtained with 
the EE method for a cubic massive neutron lattice.  It is given by 
\cite{Weise}:
\beq
\omega_{EE} - m_K = \frac{1}{m_K}\frac{2\pi a\rho}{1-a\xi\rho}
,\quad \xi=-\int d^3r \frac{C(r)}{r}  ,\label{EE}
\eeq
where  $C(r)$ is the nucleon-nucleon correlation function and
$\xi\rho$ is the inverse correlation length.
In a simple cubic lattice $\xi\rho r_0 = 1.54$, while 
realistic neutron matter wave-functions calculated in \cite{APR98}
give $\xi\rho r_0 \sim 1.23$ at densities $ \geq 2 \rho_0$.  For comparison 
$\xi\rho r_0 = 0.92$ in neutron Fermi-gas. 

At lower densities there is little difference between the EE 
results for $\xi\rho r_0 = 1.54$ and the WS or the exact results.
However, at higher densities, the $\omega_{EE}$ is larger than the 
correct result.  Note that the WS and the exact results depend upon the 
interaction range, while the $\omega_{EE}$ depends only on the 
scattering length $a$.  For larger values of $R$, same $a$,  
the exact and WS energies are lower, further below the EE result, while 
for smaller $R$ they would move up towards or even above the 
EE result.  This is to be expected, since the EE approximation assumes 
that the kaon interacts with only one nucleon at a time, which 
is valid when $r_0\gg R$.  At high densities the EE method is not 
expected to be useful, however, it does seem to be qualitatively 
good for $R \sim 0.7$ fm.
The EE results are not excessively sensitive to the value of 
$\xi\rho r_0$.  In Fig. (\ref{fig2}) we also show the results 
obtained with $\xi\rho r_0 = 1.23$ appropriate for realistic 
neutron matter.  They cross the Hartree line at 
$\rho \sim 4 \rho_0$.  

Self-energies in dilute systems can be expanded
in terms of the scattering length (the so called Galitskii's integral
equations \cite{FW}). Analogous results are obtained in chiral perturbation
theory \cite{Lee}. However, such expansions are valid only when the 
interparticle distance is much larger than the pair interaction range, 
so that only the scattering length matters.  They generally do not have the 
correct high density limit, which depends upon the shape of the 
interaction in addition to the scattering length.

The lowest order constrained variational method \cite{LOCV} used 
to treat strong correlations in nuclear matter and liquid Helium is
identical to the WS cell approximation employed here
if the healing distance is chosen as $r_0$.  These methods have 
the correct low and high density limits, and are meant to provide 
a good approximation over the entire density range.  
They are probably less accurate than the low-density 
expansions in the region where the expansions are valid.  
For example, consider the low density expansion of the WS energy 
(Eq. \ref{Lenz2}). To second order in the scattering length it
corresponds to $\xi\rho r_0=1.8 $, much larger than the 
realistic values quoted above.  Thus it is likely that at 
densities $\ll\rho_0$ the EE results for neutron matter 
shown in Fig.2 for $\xi\rho r_0=1.23$ are more 
accurate that the WS.  However, the possible error in the WS 
results at $\rho\ll\rho_0$ seems to be $< 5 \%$ by comparison.
It is also possible to calculate corrections to the Hartree 
potential at high 
densities by coupling the kaon motion to phonons. 
Their estimate \cite{PPT} at $\rho = 4 \rho_0$ is $\sim -16$~MeV 
for the square well potential with 0.7 fm radius. 
This correction decreases as $\rho$ increases.  It is larger for 
Yukawa shaped potentials (same $a$ and comparable radius) 
than the square well, however the Hartree 
potential is less attractive for the Yukawa shaped  
interaction.  

In conclusion, it seems that we can use the Hartree limit to estimate
the kaon energy in matter at densities $ \agt 4 \rho_0$ where kaon
condensation may occur.  On this basis it appears from Fig. 1 that,
for the electron potential $\mu_e(\rho)$ calculated from modern
realistic $NN$ interactions \cite{APR98}, without quark drops in
matter, kaon condensation is unlikely up to $\rho = 7 \rho_0$, which
is just above the estimated range of densities possible in neutron
stars.  With the $\mu_e(\rho)$ used in Ref. \cite{Weise}, which is
larger, the condensation could occur at $\rho \agt 6 \rho_0$.  If
dense matter has $\sim 10 \% $ protons, whose interaction with the
$K^-$ is believed to be twice as strong, the Hartree potential will be
more attractive by $\sim 10 \% $.  This together with the realistic
$\mu_e$, could reduce the condensation density to $\sim 6.5 \rho_0$.
If the $K^-N$ interaction has the Yukawa shape, or a repulsive core,
that would push the condensation density higher, and if quark drops or
hyperons reduce the value of $\mu_e(\rho)$ at higher densities as
indicated in Fig.1, kaon condensation becomes unlikely.

We acknowledge Wolfram Weise for stimulating discussions.  The work 
of VRP is partly supported by the US National Science Foundation 
via grant PHY 98-00978, the work of JC is supported
by the U. S. Department of Energy under contract  W-7405-ENG-36.


\end{document}